\documentclass[prx,superscriptaddress,twocolumn,amsmath,nofootinbib,longbibliography]{revtex4-1}

\usepackage{amsmath,graphicx,bm,wasysym,color}
\usepackage{amsmath,amsfonts,amsthm,amssymb,amsbsy,bbold,mathrsfs}

\newcommand{\pd}{\partial}

\newcommand{\yesnoNatComm}[1]{}

\newcommand{\NCarxiv}[2]{#2}

\newcommand{\yesnoMain}[1]{#1 }

\newcommand{\yesnoSupplementary}[1]{#1  }

\yesnoNatComm{\yesnoSupplementary{}}

\begin{document}

\title{Scaling and data collapse from local moments in frustrated disordered quantum spin systems}
\author{Itamar Kimchi}
\email{ikimchi@gmail.com}
\affiliation{Department of Physics, Massachusetts Institute of
Technology, Cambridge, MA 02139, USA}
\author{John P. Sheckelton}
\affiliation{Department of Chemistry, Department of Physics and Astronomy, and the Institute
for Quantum Matter, The Johns Hopkins University, Baltimore, MD 21218, USA}
\author{Tyrel M. McQueen}
\affiliation{Department of Chemistry, Department of Physics and Astronomy, and the Institute
for Quantum Matter, The Johns Hopkins University, Baltimore, MD 21218, USA}
\affiliation{Department of Materials Science and Engineering, The Johns Hopkins University, Baltimore, MD 21218, USA}
\author{Patrick A. Lee}
\affiliation{Department of Physics, Massachusetts Institute of
Technology, Cambridge, MA 02139, USA}
\date{June 27, 2018}

\yesnoMain{

\begin{abstract}
Recently measurements on various spin--1/2 quantum magnets such as H$_3$LiIr$_2$O$_6$, LiZn$_2$Mo$_3$O$_8$, ZnCu$_3$(OH)$_6$Cl$_2$ and 1T-TaS$_2$ --- all described by magnetic frustration and quenched disorder but with no other common relation --- nevertheless showed apparently universal scaling features at low temperature. In particular the heat capacity $C[H,T]$ in temperature $T$ and magnetic field $H$ exhibits $T/H$ data collapse reminiscent of scaling near a critical point. Here we propose a theory for this scaling collapse based on an emergent random-singlet regime extended to include spin-orbit coupling and antisymmetric Dzyaloshinskii-Moriya (DM) interactions. We derive the scaling $C[H,T]/T \sim H^{-\gamma} F_q[T/H]$ with $F_q[x] = x^{q}$ at small $x$, with $q \in \{0,1,2\}$ an integer exponent whose value depends on spatial symmetries. The agreement with experiments indicates that a fraction of spins form random valence bonds and that these are surrounded by a quantum paramagnetic phase. We also discuss distinct scaling for magnetization with a $q$-dependent subdominant term enforced by Maxwell's relations.
\end{abstract}
\maketitle

Heat capacity and its temperature and magnetic field dependence is a powerful tool to provide fundamental thermodynamical information on various solids  including correlated electrons in magnetic Mott insulators. It has come to our attention that recent measurements of the heat capacity of certain quantum magnets which are candidates for an exotic state of matter called quantum spin liquid\cite{Balents2010} show a power law temperature dependence and a striking one-parameter scaling and data collapse as a function of temperature $T$ and magnetic field $H$. Though the materials all appear quite different the observed scaling functions of $C[H,T]$ as power laws in $T/H$ suggest an unexpected hidden universality.

\begin{figure}[]
\includegraphics[width=\columnwidth]{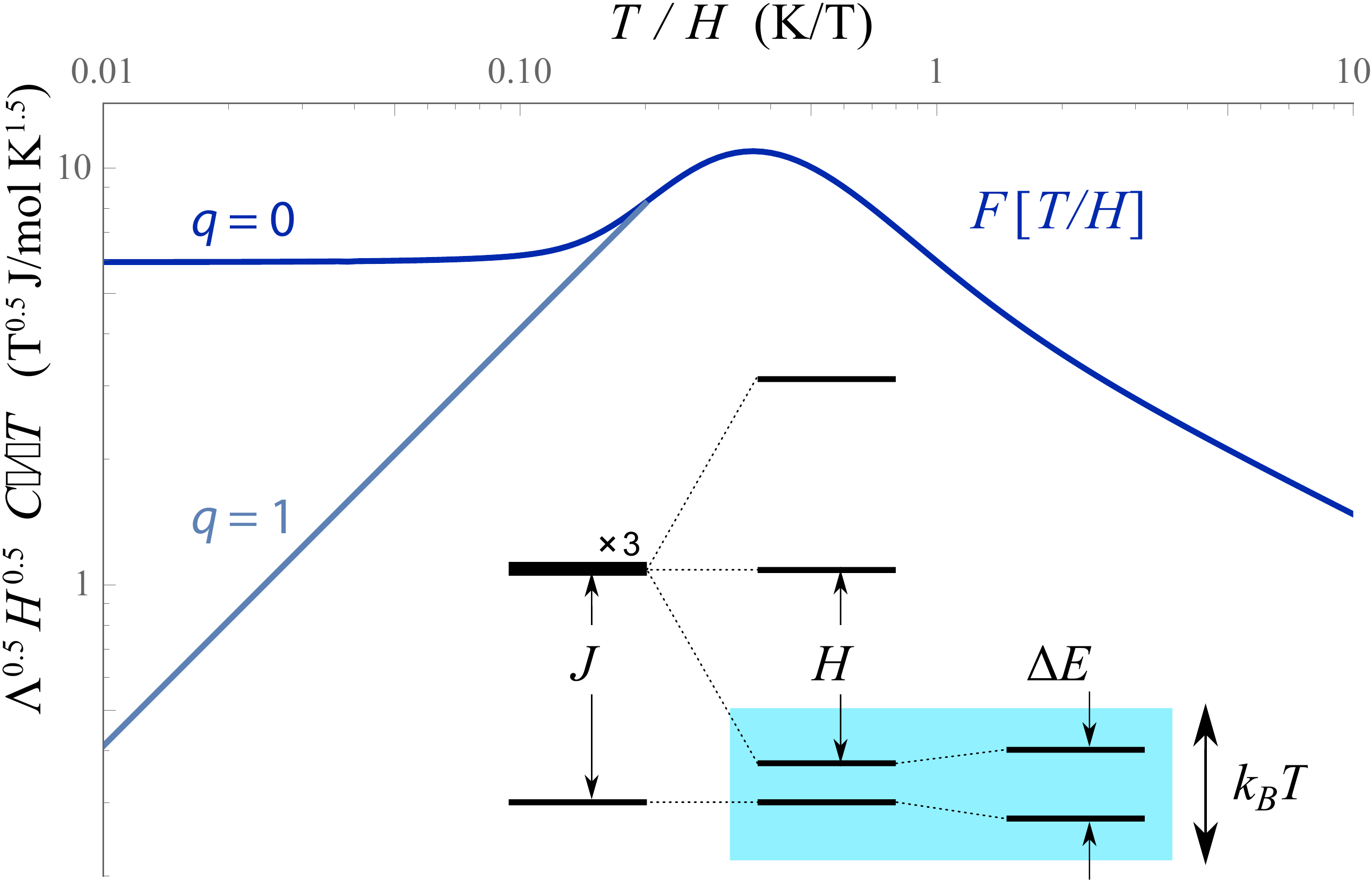}
\caption[]{ \textbf{Heat capacity scaling function $F_0[T/H]$ and its $q\,{=}\,1\,,2$ modification by level repulsion.} 
The heat capacity $C[H,T]$ for $q\,{=}\,0$ random singlets (shown with integrated energy distribution $\int P[E] = (E/\Lambda)^{1-\gamma}$ at $\gamma\,{=}\,0.5$, blue) exhibits scaling collapse in $T/H$ with $T$-linear form at $T \ll H$. 
This is easily understood (bottom inset). A spin--1/2 pair with singlet-triplet splitting $J$ acquires a resonance in magnetic fields $H \approx J$, contributing to $C[H,T]$ when $|J-H|< k_B T$. 
The scaling is modified when spin-orbit coupling and lattice symmetries combine into DM interactions with singlet-triplet mixing: the resulting level repulsion changes the resonance condition $\Delta E < k_B T$ to produce $T$-scaling with higher powers, as in the $q\,{=}\,1$ line  shown (gray).
} \label{fig:linearscaling}
\end{figure}

Power-law specific heat is a familiar consequence of the \textit{random singlet} phase seen in doped semiconductors such as Si:P, and described theoretically in 1D by Dasgupta, Ma and Fisher\cite{Ma1980, Fisher1994} and in 2D and 3D by Bhatt and Lee\cite{Lee1981,Lee1982, Sachdev1988}. In this picture the spins interact with each other via a broad distribution of antiferromagnetic exchange interactions. The spins that are most strongly coupled pair into singlets first, leaving behind spins that are further apart, eventually resulting in  a power law distribution of exchanges and a power law tail of  density of states. This picture follows from renormalization group analysis in 1D and has been demonstrated numerically in higher dimensions under a variety of conditions.
Though the original setting for the $D>1$ random-singlet phase required a dilute random network of spin--1/2 sites, without a lattice, recently a random-singlet regime has been argued to arise as a general feature of spin--1/2 lattice magnets with quantum paramagnetic ground states and random exchange energies, i.e.\ in highly-frustrated quantum magnets with quenched disorder.\cite{randommagnets}
The theory applies when the majority of spin--1/2 sites form a paramagnetic state such as a spin liquid or a valence bond crystal, and demonstrates in two different controlled limits that a small fraction of sites necessarily nucleates\footnote{As may be required by the conjectured disordered-Lieb-Schultz-Mattis restrictions\cite{randommagnets}.} and leads to a random network of spin--1/2 moments at low energies. This small subsystem may be captured by a random-singlet regime in its low temperature renormalization group flow, developing a power-law probability distribution of antiferromagnetic exchange energies $P[J] \sim J^{-\gamma}$. At a given temperature $T$, the spins with exchange $J<T$ behave as free spins giving rise to a heat capacity $C[T]\sim T^{1-\gamma}$ and spin susceptibility $\chi[T]\sim T^{-\gamma}$. In many cases the measurable response from this relatively small emergent subsystem may dominate over the response of the bulk phase. Such an emergent power-law energy distribution, associated with a relatively small portion of the spin--1/2 sites, serves as the \textit{a priori} starting point for the present work.

The power-law distribution of entropy manifests most dramatically by varying both temperature and an external magnetic field. 
Consider a given singlet bond with singlet-triplet energy splitting $J$ drawn from the distribution $P[J]$. Applying a magnetic field with Zeeman energy $H$ has no effect on the singlet energy but splits the triplet manifold.
At low temperatures $T \ll H$ this bond contributes to heat capacity only when its ground state crosses over from the singlet state to the field-polarized triplet state: the width of this resonance  $|J-H|<k_B T$ (Fig.~\ref{fig:linearscaling}) is  set by  temperature giving a  heat capacity that rises linearly in $T$. 
The distribution of bond energies enters only through the $H$-dependent coefficient: $C \sim T/H^{\gamma}$ at $T \ll H$ where $C \sim T^{1-\gamma}$ at $H=0$. (We have set the magnetic moment $g \mu_B$ to be unity.)

Given this expected scaling 
it came as a surprise that new experiments of Ref.~\onlinecite{Takagi2018} found a different scaling behavior   on the spin--1/2 magnet H$_3$LiIr$_2$O$_6$, a member of the family of so-called Kitaev honeycomb iridates\cite{Kee2016,Knolle2018}. 
Both magnetic frustration and disorder are likely ingredients in this compound: exchanges between Ir$^{4+}$ effective spin--1/2 moments may vary randomly depending on the positions of mobile hydrogen ions, and the frustration expected from iridium's strong spin-orbit coupling is evidently manifest in the lack of any ordering transition down to at least 0.05 K, less than a percent of the 100 K exchange energy scale. The bulk of the sites form a quantum paramagnetic phase, such as a spin liquid. The remaining fraction of sites was found to contribute a power-law heat capacity $C \sim T^{1/2}$ with an appropriate small coefficient as expected from a random-singlet regime; but when a magnetic field was applied, instead of $T$-linear heat capacity, clear \textit{quadratic}  scaling $C \sim T^2 / H^{3/2}$ was observed for $T \ll H$. Furthermore, as shown in the inset of Fig.~4(a) in Ref.~\onlinecite{Takagi2018},
the data are found to collapse to a single scaling curve of the form $C[H,T]/T \ \sim \ \ H^{-\gamma} \ F[T/H]$ where $\gamma=0.5$.

\begin{figure}[]
\includegraphics[width=0.75\columnwidth]{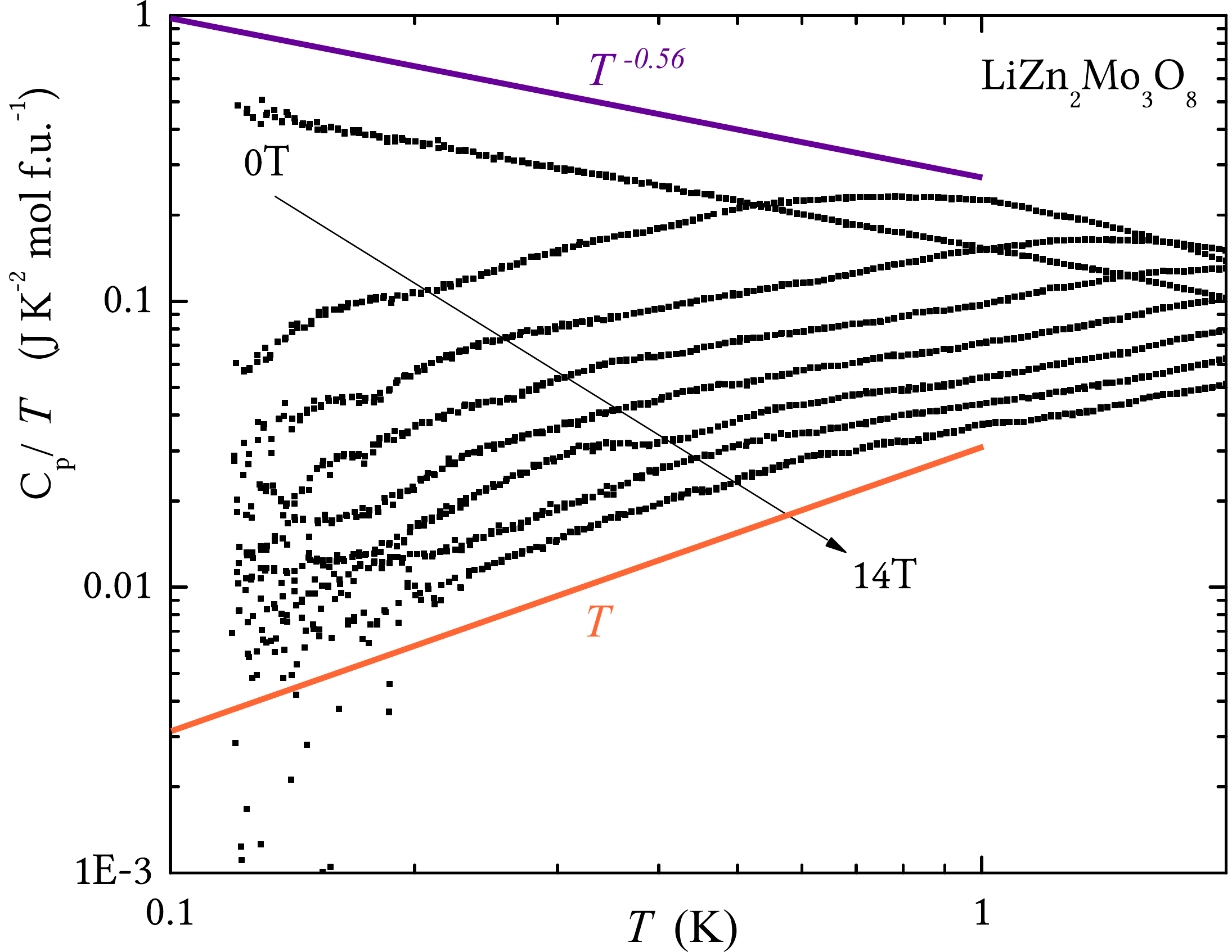}
\caption[]{ \textbf{Power law heat capacity in LiZn$_2$Mo$_3$O$_8$ under various magnetic fields.} At zero field the heat capacity shows a non-integer power law $C/T \sim T^{-0.56}$ (purple line), changing under various magnetic fields to functional forms with quadratic $C\sim T^2$ behavior (orange line) at low temperature.
} \label{fig:lzmopowerlaws}
\end{figure}

\begin{figure}[]
\includegraphics[width=\columnwidth]{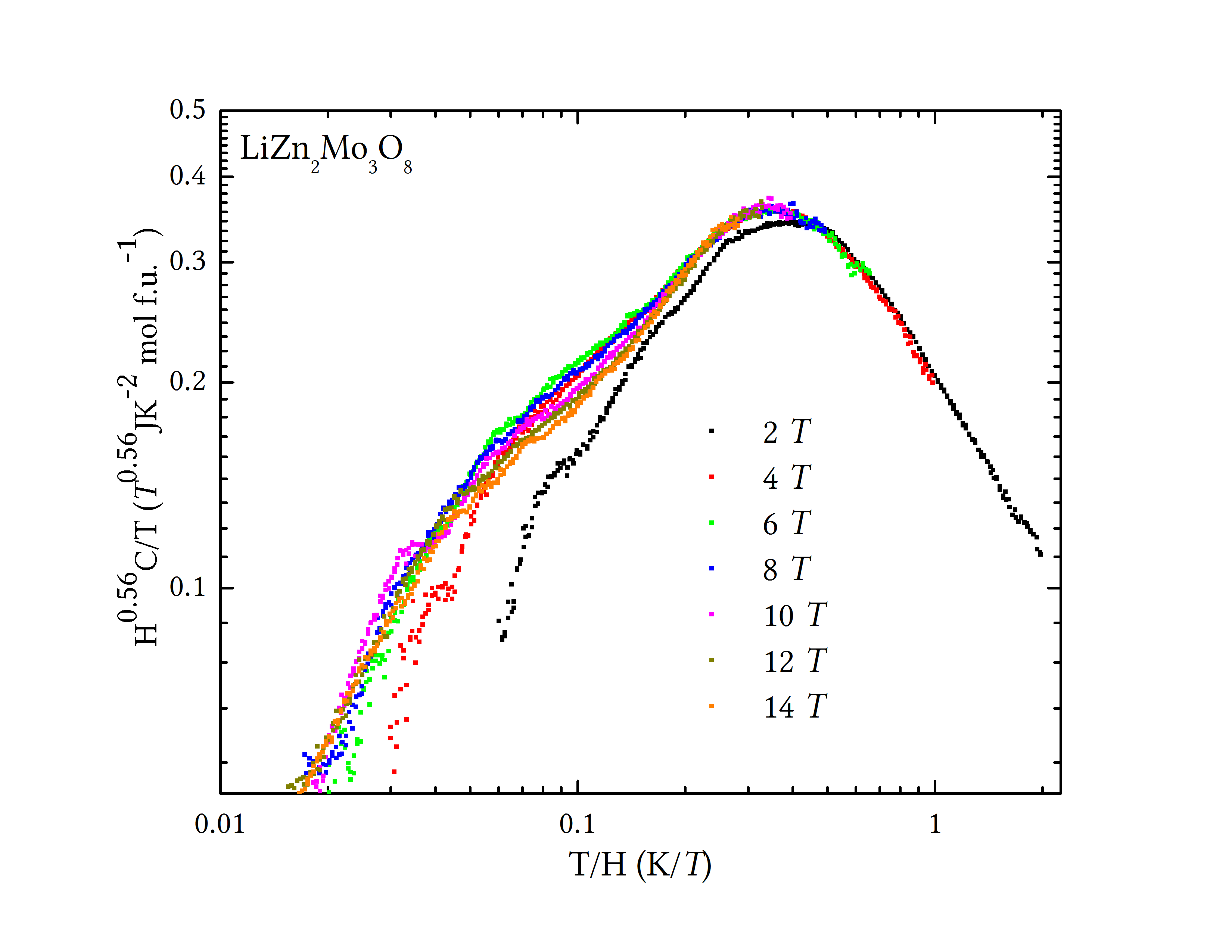}
\caption[]{ \textbf{Data collapse in LiZn$_2$Mo$_3$O$_8$.} 
Data from Fig.~\ref{fig:lzmopowerlaws} rescaled by the scaling ansatz  $H^{0.56} C/T$. 
The data collapses to a function of the single variable $T/H$, asymptoting as $(T/H)$ for $T{\ll} H$ and $(T/H)^{-0.56}$ for $T {\gg} H$, consistent with the $q=1$ scaling theory Eq.~\ref{eq:scaling} and Fig.~\ref{fig:linearscaling}. 
} \label{fig:lzmo}
\end{figure}

The work of Ref.~\onlinecite{Takagi2018} reminds us of a system we had studied earlier, LiZn$_2$Mo$_3$O$_8$. This is a layered magnet where 2/3 of the spin disappears into singlets, leaving behind 1/3 of the spins which behave almost as free spins  \cite{McQueen2012,McQueen2014,Broholm2014}. The mechanism for this behavior is not well understood, since various theoretical proposals\cite{Lee2013a,Kim2016,Lee2018} all require some form of short range tripling of the unit cell size, which has been searched for and not found\cite{Broholm2014}. 
However on symmetry grounds we again expect the prevalent non-magnetic Li/Zn site mixing disorder\cite{McQueen2012} to generate bond randomness, whose competition with singlets requires\cite{randommagnets} low energy spin excitations to appear. 
Here we focus our attention on the fate of these remaining spins at low temperatures. It turns out that  independent of Ref.~\onlinecite{Takagi2018}, we had recognized that our previously unpublished heat capacity data (Fig.~\ref{fig:lzmopowerlaws})   also show clear data collapse with $T^2$ scaling (Fig.~\ref{fig:lzmo}). 
Similar data collapse, though with a smaller accessible $T/H$ window, is seen in previous heat capacity measurements 
from Ref.~\onlinecite{Lee2007} for synthetic herbertsmithite ZnCu$_3$(OH)$_6$Cl$_2$ (Fig.~\ref{fig:herbertsmithite}). 
These materials, and related ones which we will discuss further below --- 1T-TaS$_2$, YbMgGaO$_4$, YbZnGaO$_4$ and Ba$_2$YMoO$_6$ ---  differ in their spin--1/2 lattices, magnetic Hamiltonians, spatial symmetries, and sources of randomness; but their lack of magnetic order down to the lowest temperatures measured, the presence of some randomness, and the observed power-law scaling laws consistent with a contribution from a fraction of sites, taken together call for a unified theoretical framework.

\begin{figure}[]
\includegraphics[width=\columnwidth]{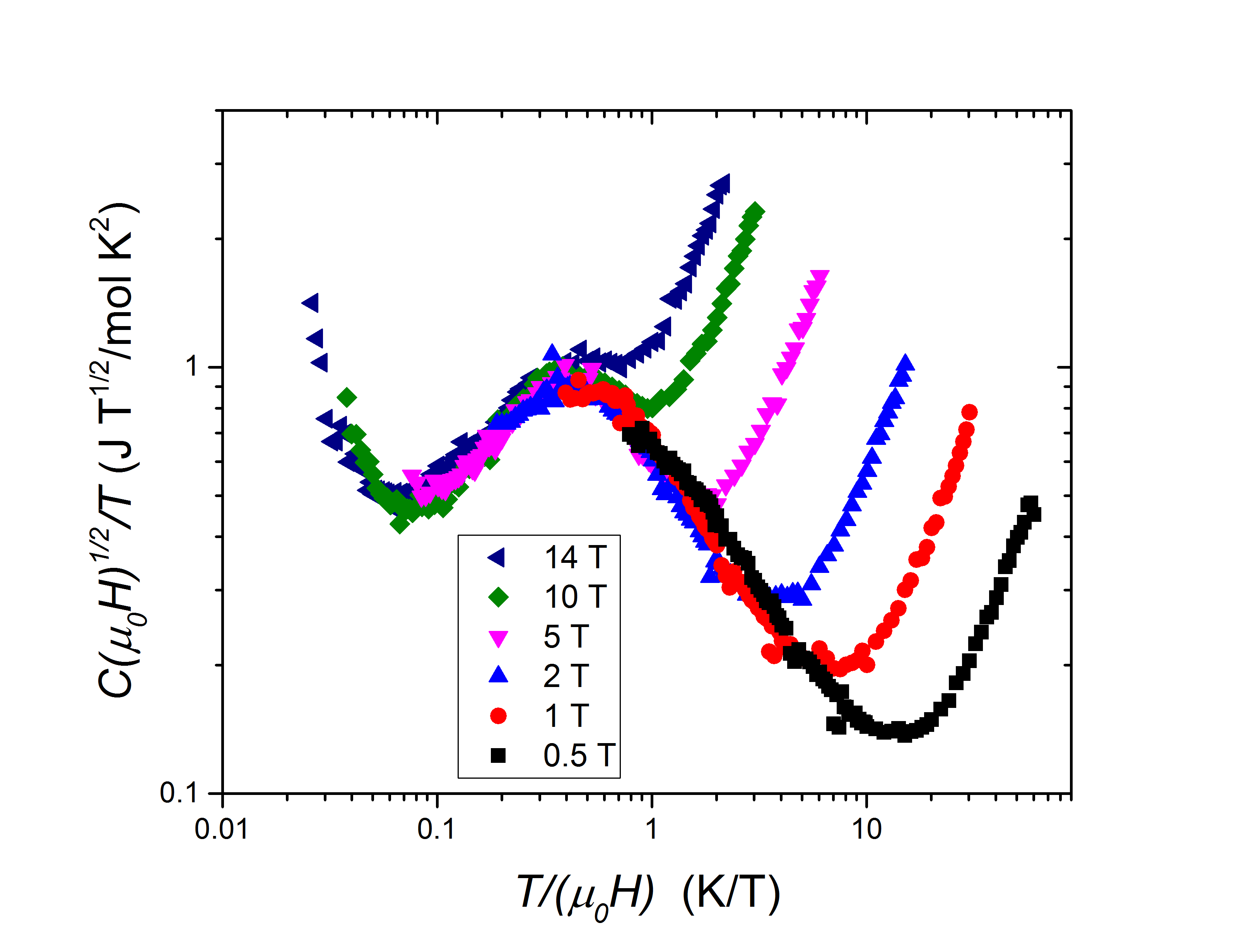}
\caption[]{ \textbf{Data collapse in 
synthetic herbertsmithite ZnCu$_3$(OH)$_6$Cl$_2$.} 
Herbertsmithite heat capacity data from Ref.~\onlinecite{Lee2007},   here replotted using the scaling ansatz  $H^{0.5} C/T$ (per formula unit of ZnCu$_3$(OH)$_6$Cl$_2$) to show data collapse. The phonon contribution has not been subtracted, which accounts for the upturn for $T/H \gtrsim 1$. 
Away from the upturn at low $T/H \lesssim 0.05$, which is due to the nuclear Schottky contribution, the data collapses with a peak consistent with the scaling function Eq.~\ref{eq:scaling} and Fig.~\ref{fig:linearscaling}. 
} \label{fig:herbertsmithite}
\end{figure}

In this work we argue that the missing factor of $T/H$ needed to explain the more recent data is captured by an extension of the random-singlet theory that includes spin-orbit coupling (SOC) and its interplay with spatial symmetries, in particular through antisymmetric Dzyaloshinskii-Moriya (DM) spin exchanges. 
For the sake of continuity we will denote the theory as \textit{random-singlets}, even though the resulting theory describes a configuration of non-degenerate valence bonds that are no longer singlets under spin rotation.  
We will show how this produces three possibilities for $T$ dependence of heat capacity in a magnetic field, associated with data collapse in $T/H$ and a scaling exponent that can take one of three integer values. 

\yesnoNatComm{\subsection{Results}\subsubsection{Scaling and data collapse}}

The derivation of the $T$-linear scaling shown in Fig.~\ref{fig:linearscaling} assumed that the downshifted  triplet state crosses the singlet state freely.  This assumption no longer holds in the presence of spin orbit coupling. The triplet will in general be split but in a magnetic field one level will still move down to approach the ground state with a modified $g$ factor.  Any off-diagonal matrix element $D$ will mix these states, leading to level repulsion and resulting in an additional, independent, constraint $D<T$ supplementing the diagonal constraint $|J-|H||<T$. It is important to note spin-orbit coupling is a necessary but not sufficient ingredient for the level repulsion: spin-orbit coupling modifies the magnetic exchanges in two distinct ways depending on spatial symmetries.
With an inversion center at the bond midpoint, or certain other combinations of symmetries, the matrix $J_{i j}^{\alpha \beta}$ of the spin exchange $J_{i j}^{\alpha \beta}S^\alpha_i S^\beta_j$  is required to be symmetric. The two-spin singlet state is odd under the inversion symmetry and does not mix with the triplet manifold. However without these symmetries, the spin exchange matrix gains an antisymmetric contribution to $J_{i j}^{\alpha \beta}$ which is conventionally described by the DM term $\vec D \cdot (\vec S_i \times \vec S_j) $. The magnitude of the DM vector $\vec D$ is linear in the strength of SOC for weak SOC. By breaking inversion symmetry $i \leftrightarrow j$ it mixes the singlet and the triplet and produces the desired level repulsion.

The preceding argument may be considered in more detail. While the zero-field splitting is given simply by $(1/2)\sqrt{J^2 + |D|^2 }$, 
in a magnetic field this splitting becomes (Supplementary \NCarxiv{Note 1}{\ref{app:twospins}}) 
\begin{align} \Delta E = \frac{1}{2\sqrt{2}}\sqrt{2(|H|-J)^2 + D_1^2 + D_2^2} \label{eq:split}\end{align}
where $D_1,D_2$ are the components of the DM vector $\vec D$ that lie perpendicular to the magnetic field $\vec H$.
It is clear that each component of the DM vector $\vec D$ that enters into the resonance condition Eq.~\ref{eq:split} contributes a factor of $T/D$ to the scaling of specific heat at low $T$  (Supplementary \NCarxiv{Note 3}{\ref{app:heatcapacity}}). Furthermore, as long as the ratio of DM interactions to symmetric exchange interactions remains roughly fixed in the RG flow, these factors can be replaced by $T/H$. We show that this  is indeed the case within the strong-disorder-RG renormalization step (Supplementary \NCarxiv{Note 2}{\ref{app:SDRG}}). The details of the scaling functions can depend on crystal symmetry as well as the dimensionality of the magnetic lattice (Supplementary \NCarxiv{Note 4}{\ref{app:results}}). 
We thus find three possibilities: the linear scaling $C[T] \sim  T /H^\gamma$ may stay the same or gain a factor linear or quadratic in $T/H$. 

This result can be captured by a general scaling function $F_q$ for the density of states as measured by heat capacity, 
\begin{align}
\label{eq:scaling}
\frac{C[H,T]}{T} \ \sim& \ \ \frac{1}{H^{\gamma}} \ F_q\left[\frac{T}{H}\right]
\\
 F_q[X] \ \sim& \  \left\{ \begin{array}{cc}
 X^{q} & \quad X \ll 1\\
 X^{-\gamma}\left(1 + \frac{c_0}{X^2}\right)& \quad X \gg 1 \end{array}
 \right.
 \nonumber
\end{align}
The sub-dominant  scaling term at large $T/H$ must be added in order to satisfy the Maxwell relation between entropy and magnetization. A related subdominant term must be added to the scaling of magnetization at $T\ll H$ yielding $M/H\sim H^{-\gamma}\left(1-m_1 (T/H)^{q+2}\right)$
(for $m_1$ and further discussion see Supplementary~\NCarxiv{Note 5}{\ref{app:magnetization}}).
Here the coefficient $c_0$ is given by $c_0 = ((1+\gamma)\gamma/2) /(C_{H=0}/T \chi_{H=0})$ when the zero-field susceptibility is $\chi_{H=0} \sim T^{-\gamma}$.
This scaling form shows the familiar non-universal exponent $0 \lesssim \gamma \lesssim 1$ that characterizes the random-singlet distribution, but in addition  a new integer index $q$ appears that takes one of the three values $\{0,1,2\}$ for the three cases described above; its choice depends on the interplay of SOC with spatial symmetries as we will discuss below. 

Two different values of $q$ are necessary to capture the observed experimental scaling we are aware of so far.  We have already discussed the quadratic scaling ($q=1$) observed for H$_3$LiIr$_2$O$_6$ and LiZn$_2$Mo$_3$O$_8$.
The $q=1$ scaling in these layered magnets may be understood as arising from a DM vector that is forced to point perpendicular to the lattice plane by an approximate reflection symmetry across the plane that emerges during the course of RG flow: in each strong-disorder RG step the new spin exchanges are generated based on the pattern of valence bonds that have been integrated out at higher energies, and for a 2D lattice any such configuration of valence bonds preserves the mirror reflection across the plane, leading (Supplementary \NCarxiv{Note 4}{\ref{app:results}}) to a single-component DM vector and $q=1$. 

In Fig.~\ref{fig:herbertsmithite} we show the data collapse for the well studied spin--1/2 kagome lattice material ZnCu$_3$(OH)$_6$Cl$_2$. It is known that approximately 15\% of the Zn sites which are off the kagome planes are replaced by Cu, forming $S$=1/2 local moments\cite{Lee2016}. Furthermore, there is significant DM coupling. We should mention two caveats. First the nuclear contribution makes the low temperature spin contribution inaccessible and second, the bulk spin gap is estimated to be 0.7 meV, so that the high field data may have some bulk contribution due to the closing of the spin gap. With these reservations, the data collapse shown in Fig.~\ref{fig:herbertsmithite} may also be consistent with the $q=1$ case.

Next we consider the layered material 1T-TaS$_2$, where a charge density wave (CDW)   transition at intermediate temperatures leads to a cluster Mott insulator with one spin--1/2 per 13-site unit cell arranged into a triangular lattice, with phenomenology consistent with a spin liquid state\cite{Lee2017}. A small fraction of the spins (less than than a few \% per cluster) were recently observed to produce a low temperature $T$-linear term in the heat capacity with a coefficient that decreases monotonically with a magnetic field (Ref.~\onlinecite{Kanigel2017} Fig.~4) roughly as $C\sim H^{-2/3} T$. This is consistent with the argument above for linear scaling ($q=0$) for 1T-TaS$_2$ due to its high crystal symmetry, preserved even in the CDW state, that forbids DM interactions from being generated for not only nearest neighbor and second neighbor bonds but further neighbor bonds as well. 
A replot of the $C[H,T]$ two-parameter data (Y. Dagan and I. Silber, personal communication) indeed shows data collapse  consistent with Eq.~\ref{eq:scaling} with $q=0$.

It will be interesting to test the data collapse scaling form for other frustrated spin--1/2 quantum magnets.
Other layered compounds such as YbMgGaO$_4$ and YbZnGaO$_4$, both with effective spin--1/2 moments from Yb arranged on a triangular lattice and with intrinsic Mg/Ga and Zn/Ga charge disorder\cite{srep,Gegenwart2017} show clear $C\sim T^{1-\gamma}$ scaling in zero field measurements\cite{srep, Li2016, Wen2018}; 
but the $T \ll H \ll J_0$ scaling limit, given the small lattice magnetic exchange energy $J_0 \sim$ few K,  is not yet clear. The case $q=2$ is expected to arise most naturally in magnets with fully 3D magnetic lattices such as Ba$_2$YMoO$_6$\cite{Bos2010}, which we leave for future work.

Finally, we note that the appearance of scaling and data collapse sheds light on two interrelated subsystems of the quantum magnet. 
The first subsystem is formed by the local moments that contribute to the heat capacity scaling. 
When the material disorder does not directly change the sites of its spin--1/2 lattice but rather impacts it (weakly or strongly) only through bond randomness in the magnetic exchanges --- likely the case for all the materials discussed above except synthetic herbertsmithite --- then  the local moment subsystem is  \textit{emergent} via an unusual RG flow\cite{randommagnets}. The quantum critical scaling functions which can be exhibited by this subsystem are interesting both in their own right and as a signature of a coexisting quantum paramagnet state for the second subsystem, consisting of the remaining spins. This coexisting quantum paramagnetic phase must involve valence bonds which may be frozen, possibly with a relic of valence-bond-solid order, or resonating, as in a quantum spin liquid and associated topological order.  The interplay of the quantum paramagnet with the local moments that produce the $T/H$ scaling merits further exploration. 

\textit{Acknowledgments.} 
IK thanks Adam Nahum and T. Senthil for discussions and instrumental related work, and Martin Mourigal for relevant discussions. TMM and JPS acknowledge Oleg Tchernyshyov for
useful discussions. We thank Joel Helton and Young Lee for discussions and for permission to show their data collapse plot Fig.~\ref{fig:herbertsmithite}. We thank Hidenori Takagi for discussions. We also thank Yoram Dagan and Itai Silber for sharing their data and plots for 1T-TaS$_2$.  IK acknowledges support by the Pappalardo fellowship at MIT. 
The work at IQM was supported by the U.S.
Department of Energy, office of Basic Energy Sciences, Division of
Materials Sciences and Engineering under Grant No. DEFG02-08ER46544. TMM
acknowledges support of the David and Lucile Packard foundation. 
PAL acknowledges the support of DOE Grant No.\ DE-FG02-03-ER46076.

\yesnoNatComm{
\textit{Author Contributions.} 
J.P.S. and T.M.M. contributed to the synthesis and measurements of LiZn$_2$Mo$_3$O$_8$. I.K., T.M.M., and P.A.L. contributed to the conception and analysis of the theoretical problem and to the writing of the paper.

The authors declare no competing interests.

\textit{Data Availability.} 
The data generated in this study are available from the authors on reasonable request.
}

}

\yesnoSupplementary{



\NCarxiv{ }{\appendix}
\section*{Supplementary Information}

This Supplementary contains important details for the calculations described in the main text. 
It is structured as follows. We  begin by considering the effect of DM interactions on modifying the two-spin singlet state, as well as the competition of these effects with an applied magnetic field. We then consider the RG flow under strong disorder RG, as appropriate for random singlets, in the presence of DM interactions. Then we compute the expected scaling phenomenology and discuss the various cases for $q$. Finally we relate the scaling forms for various observables including heat capacity, NMR lifetimes and magnetization.

The starting point for the calculations in this Supplementary is a picture of emergent couplings with the topology of a random network, a distribution with a long power-law tail, and an associated strong-disorder RG flow. To understand the emergence of such self-similar power laws out of a microscopic lattice model with merely flat disorder distributions, we refer the interested reader to the RG flows discussed in Ref.~\onlinecite{randommagnets}. A picture for how the RG flow is initiated out of the UV lattice scale for one simple class of microscopic models is depicted in Fig.~\ref{fig:vbs}.

\begin{figure}[]
\includegraphics[width=0.75\columnwidth]{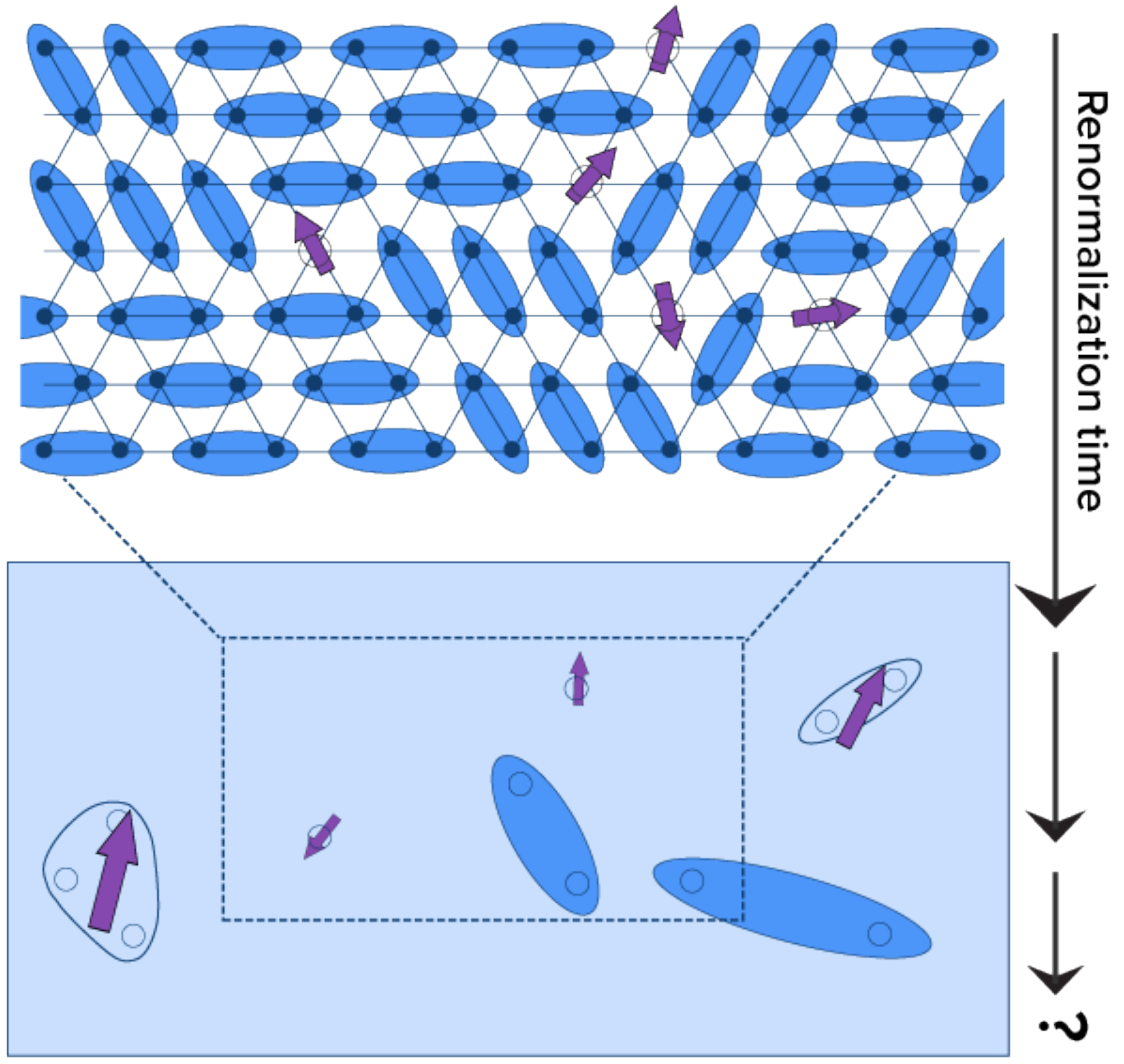}
\caption[]{ \textbf{RG flow from a class of weakly-disordered lattice models into a random network with power law distributions.}
Disordering short-ranged valence bonds is found, under certain conditions and even for weak randomness, to necessarily lead to defects that carry protected spin-1/2 moments. The resulting spin moments form a random network with power-law distributions, an appropriate setting for strong-disorder RG. Regardless of the ultimate fixed point of the RG, which in 2D and 3D is unknown, the system is described by emergent power laws across some low energy regime associated with the RG flow. 
 Adapted from Ref.~\onlinecite{randommagnets}, CC-BY-3.0.
} \label{fig:vbs}
\end{figure}

\subsection{Two spins with DM interactions}
\label{app:twospins}
To see the effects of DM interactions on the valence bond states, consider the following Hamiltonian for two spins $i,j$,
\begin{align} \mathcal{H}_{i j}= J (\vec S_i \cdot \vec S_j) + \vec D \cdot (\vec S_i \times \vec S_j) - \vec H \cdot (\vec S_i + \vec S_j)\end{align}
where in addition to a general DM term and magnetic field, for simplicity here the symmetric spin interactions are taken to consist of pure Heisenberg exchange; below we discuss generic SO(3) breaking exchanges.  
Defining the total spin operator $\vec S_{\text{tot}} \equiv  \vec S_i + \vec S_j$, the Hamiltonian up to an overall energy shift may be written as
\begin{align} \mathcal{H}_{i j}= \frac{J}{2} (\vec S_{\text{tot}})^2  + \frac{|D|}{2} (T_D + T_D^\dagger) - \vec H \cdot \vec S_{\text{tot}} 
\end{align}
where the DM term mixes the singlet with the member of the triplet manifold that has zero total spin moment along the DM vector,
\begin{align}
T_D \equiv i \ |\text{singlet} \rangle \langle \text{triplet; }  (\vec S_{\text{tot}} \cdot \vec D =0)\ | 
\end{align}

At zero field $H=0$, the ground state is a mixture of the spin-singlet state and this $\vec S_{\text{tot}} \cdot \vec D =0$ member of the triplet manifold, set by the Hamiltonian matrix $2\mathcal{H}_{H{=}0}=J \sigma^3 + |D| \sigma^2$ acting on the basis of these two states. The ground state is nondegenerate; for small $D/J$, it is modified from the singlet state only perturbatively in  $D/J$.

For larger fields near resonance $H \approx J$ it is useful to project the Hamiltonian to the low energy space spanned by the singlet and the member of the triplet with maximal total magnetization along the magnetic field direction. In this 2-level subspace, the Hamiltonian, again up to a constant, is 
\begin{align} \mathcal{H}_{H{\approx}J}= \frac{1}{2}(|H|-J) \sigma^3   +\frac{1}{2\sqrt{2}} \left( D_1 \sigma^2 - D_2 \sigma^1 \right)  \label{eq:Hh2} \end{align}
where $\sigma$ are Pauli matrices acting on the (singlet, total-spin-up-triplet) basis, and $D_1, D_2$ are the components of the DM vector perpendicular to the magnetic field axis.

Indeed this equation, with modified parameters, describes the general scenario reached by adding generic spin-orbit-coupled interactions to the symmetric spin exchange matrix $J_{i j}^{\mu \nu} S_i^\mu S_j^\nu$.
Without the DM term the singlet state remains an eigenstate and we shall assume the symmetric exchanges are sufficiently antiferromagnetic so that the singlet is the ground state of $J_{i j}^{\mu \nu} S_i^\mu S_j^\nu$. The triplet-manifold eigenstate brought down by the field is modified. The resulting Hamiltonian is described by Eq.~\ref{eq:Hh2} with $J$ interpreted as an effective exchange parameter that may depend on the direction of $\vec H$, and $D_1,D_2$ interpreted as two particular components of the DM vector.

\subsection{Strong-disorder RG step: integrating out two spins with DM interactions}
\label{app:SDRG}

The strong disorder RG (SDRG) step entails integrating out a pair of spins and considering the renormalization of interaction for every other remaining pair. We perform it analytically, within a controlled hierarchy of parameters, to establish a recursion condition that must be satisfied by any fixed-point distribution. The recursion condition ensures (1) that the Heisenberg SDRG is not modified by the presence of DM interactions, and (2) that the flow of DM interactions preserves the separation of scales discussed in the main text and used in the derivation of the specific heat scaling (Sec.\ref{app:heatcapacity}).

In one-dimensional systems SDRG is possible to perform analytically in certain cases\cite{Fisher1994}. 1D systems are special for a variety of reasons, most importantly here since integrating out two neighboring spins in a spin chain results in a system that is still exactly described as a spin chain. Numerical implementations of SDRG in higher dimensional systems, using various approximations, generally find that the SDRG assumptions become uncontrolled and the fate of the ultimate fixed point remains controversial\cite{Lee1982,Fisher2000,Igloi2003,Troyer2012}.

 Here we avoid such unresolved questions about the fixed point of $d>1$ SDRG by restricting ourselves to a particular narrow question: how does the presence of DM interactions change the SDRG? We are able to answer this question rigorously by performing a single SDRG step analytically and noting two observations about the resulting recursion relation. First, we find that weak DM interaction does not enter the renormalization of the symmetric (e.g.\ Heisenberg) exchanges. The SDRG for Heisenberg exchange therefore proceeds identically as in the case without DM interaction. Second, we find that the renormalization of DM interactions preserves a parametric separation of scales between the DM and symmetric exchanges on each bond. The relative scale of the DM interactions therefore does not change and in particular also the control parameter for the first result is preserved under RG. In this precise sense, the DM interactions are merely spectators to the standard SDRG, which proceeds as usual.

\textbf{Computation.}
To perform the single SDRG step it suffices to look at each cluster of four spins containing the strong pair. 
We assume weak DM interactions; this is the case for weak spin-orbit coupling as well as for strong spin-orbit coupling with some approximate microscopic or emergent inversion or mirror symmetries. As the starting point for an SDRG step we must postulate that the quantum state of a portion of the system, namely the fraction of spins that participate in the random-singlet-type regime, is described by a power law distribution of exchange energies. 
Computing the SDRG step in the presence of weak DM interactions, we find that in 1D the DM and Heisenberg terms scale identically, while in 2D they scale at the same order with some distinctions in detail. The resulting \textit{random-singlet} state has each spin paired with another into a nondegenerate frozen state, that differs from the spin-singlet state by the addition of a spin quadrupole (nematic) moment that preserves time-reversal, as is anyway required by the lack of spin rotation symmetry.

Consider a pair (1,2) of strongly coupled spins, with Heisenberg exchange $J_{1 2}$ as well as DM exchange $D_{1 2}$, and consider weak Heisenberg as well as DM interactions between spins 1,2 and two additional spins 3,4. 
The ground state of $H_{1 2} = J_{1 2} H_{1 2}^\text{Heis} + D_{1 2} H_{1 2}^\text{DM}$ is no longer a spin-singlet under spin rotations, but nevertheless is a unique state. 
The low energy manifold has spins 1,2 in the ground state of $H_{1 2}$, while the bare interactions $H_{3 4}^0$ gain a renormalized contribution by virtual excitations of the (1,2) excited states, as $H_{3 4}^0 \rightarrow H_{3 4}^0+ \Delta H_{3 4}$. 
Here and below we take $J_{1 2}$ to be larger than all other Hamiltonian parameters i.e.\ we work within lowest order perturbation theory in  $1/J_{1 2}$. 
 Integrating out spins 1,2, we find the renormalized interaction $\Delta H_{3 4}$ which can be expressed with coefficients $J_{3 4}$, $D_{3 4}$ as $\Delta H_{3 4} = J_{3 4} H_{3 4}^\text{Heis} + D_{3 4} H_{3 4}^\text{DM}$. 
In writing the results below we specialize to the case where all DM vectors share the same orientation. The resulting expressions are directly applicable to the $q=1$ case, e.g.\ a 2D layered system with effective reflection symmetry across each 2D lattice plane. This is the relevant analysis for the experimental data shown in the figures of the main text, as well as for the data of Ref.~\onlinecite{Takagi2018}. 

The renormalized contributions to the Heisenberg and DM interactions are given by the following expressions. (We write the DM vector on an oriented bond $(i,j)$ as a pseudoscalar $D_{i j}$.) The symmetric splitting remains exactly unchanged by the presence of DM interactions,
\begin{align}
& J_{3 4} \ = \  \frac{(J_{1 4} - J_{2 4})(J_{2 3} - J_{1 3})}{2 J_{1 2}}
\end{align}
Only the emergent DM interaction is modified,
\begin{align}
D_{3 4} \  =  &
\ \ \ \frac{D_{1 2}\,  (J_{1 4} J_{2 3}  - J_{1 3} J_{2 4}) }{2 J_{1 2}^2} \nonumber
\\ & -
\frac{(D_{1 3} - D_{2 3})(J_{1 4} - J_{2 4}) }{2 J_{1 2}} \nonumber
\\  & -
\frac{(D_{1 4} - D_{2 4})(J_{1 3} - J_{2 3}) }{2 J_{1 2}}
\end{align}
Importantly, if we assume a parametric separation of scale between the DM interaction and the symmetric splitting on each bond, this small parameter appears in all three terms above, and thus the resulting DM interaction ratio $D_{3 4}/J_{3 4}$ appears with precisely the same small parameter.

The conclusion arising from this result, which remains true for arbitrary orientations of DM interactions, is thus as follows. 
A distribution of symmetric (e.g.\ Heisenberg) and antisymmetric (DM) interactions, characterized by a parameter denoting the ratio of strengths of antisymmetric and symmetric interactions on every bond, shows an SDRG evolution that obeys two important properties.
\begin{enumerate}
\item The distribution of symmetric exchanges ($J$'s) evolves in exactly the same way regardless of the presence or absence of DM interactions. The DM interactions are spectators and do not modify the SDRG evolution of Heisenberg exchanges.
\item The distribution of DM interactions evolves in a way that does depend on the distribution of $J$'s but nevertheless preserves the condition of separation of scales, namely the ratio of DM and $J$ energy scales is preserved. The overall scale of the distribution of $|D_{i j}/J_{i j}|$ remains fixed under SDRG.
\end{enumerate}
This result allows us to take the known results on SDRG with Heisenberg terms and apply them directly to the present case with additional DM interactions.

\subsection{Derivation of scaling in a magnetic field}
\label{app:heatcapacity}
Here we discuss in detail the impact of DM interactions on the scaling relation for the density of states, as observable by heat capacity, in a magnetic field. 

The entropy distribution at low energies is set by the distribution of couplings that emerges under RG flow.
Consider then the distribution of bonds picked by the strong-disorder RG.
It defines the (highly correlated) distributions of three energy parameters:
(1) the symmetric splitting $J$; 
(2) the antisymmetric splitting set by the components $D_i$ of the DM vector $\vec D$; 
(3) the resulting true splitting between the lowest state and the second state.
Let us denote the distributions of these parameters by $P_1, P_2, P_3$ respectively. 
The Hamiltonian matrix elements for the lowest two states are given above, and the resulting splittings are as follows. 
With no magnetic field, this splitting is
\begin{align} \Delta E [H=0] = \frac{1}{2}\sqrt{J^2 + |D|^2 } \end{align}
In a magnetic field, this splitting (see Eq.~\ref{eq:Hh2}) becomes
\begin{align} \Delta E = \frac{1}{2\sqrt{2}}\sqrt{2(|H|-J)^2 + D_1^2 + D_2^2} \end{align}
where $D_1,D_2$ are the components of the DM vector that lie perpendicular to the magnetic field $\vec H$.
The entropy distribution seen directly by $C[T]$  at zero field  is the distribution of the  splittings $P_3[E]$.
Self consistency within the strong-disorder RG framework requires a broad tail for $P_3[E]$, generally power law $P_3[E] = E^{-\gamma}$, with $\gamma$ defined by the specific heat exponent at zero field, 
\[C[H=0][T] \sim T^{1-\gamma}.\]
 Now let us add a magnetic field and consider the scaling of specific heat at temperatures far below the Zeeman energy, $T \ll H$. 

First recall the case of no DM interactions, $D = 0$.
Here the field picks out the distribution $P_3[H]$, ie $P_3$ at energy $\Delta E = H$. Now we may linearize $P_3$ for energy splittings E near $\Delta E = H$. It has an analytic polynomial expansion, which to first order may be written as 
$P_3[E] = P_3[H] + c_1 (E-H)$. The specific heat picks out all states with energy $\Delta E < T$. Due to the linearization, $C[H,T] \approx T P_3[H]$ .

Now consider the case of small but nonzero DM interactions. Let us count the states with excitation energy $\Delta E < T$.
$J$ is set by the distribution $P_1[J-H] = P_1[H] + c_1 (J-H)$. 
The conditional distribution for each component $D_i$ of the DM vector $\vec D$, conditioned on being given a bond with a particular Heisenberg exchange $J \approx H$, may be taken to be approximately Gaussian with some width set by this $J\approx H$. 
To reach a resonance with $\Delta E < T$ in the regime where $T \ll H$, we must have  $D_i \ll H$ and thus the distribution of $D_i$ of interest here is uniform and may be approximated by its value at the origin, $P_2[D_i] \approx P_2[0]$.
 Since the width of the distribution $P_2$ is proportional to $J\approx H$, normalization sets its overall amplitude by a factor proportional to $1/J \approx 1/H$, giving $P_2[D_i] \approx P_2[0] = c_2 / H$. The number of states with $\Delta E < T$ is then given by $T P_1[H] \sim T /H^{\gamma}$ times a factor of $T/H$ for each relevant component $D_i$.

\subsection{Scaling of specific heat in a magnetic field}
\label{app:results}
 
Due to quantum mechanical level repulsion, a low excitation energy $\Delta E < T$ entails conditions on the amplitude of three parameters: $||H|-J| < T$, $|D_1| < T$, and $|D_2| < T$. We must now consider three distinct physical scenarios, depending on the crystal symmetries or the effective approximate symmetries that emerge in SDRG, as well as potentially depending on the direction of the magnetic field. Recall that the scaling relation is $ C[H,T] \sim T^{1-\gamma}$  for $H \ll T$, and is
\begin{equation} C[H,T] \sim \frac{T^{q+1}}{H^{q+\gamma}} \end{equation}
for $T \ll H$.

 Case (1): $\vec D \times \vec H = 0$. This is the case if symmetries constrain the DM vector to lie along a particular axis, and the magnetic field is taken to lie along the same axis. Then the specific heat scaling reduces to the case with no DM terms, and we find  $q=0$.
 
 Case (2): only one component of $\vec D$ perpendicular to $\vec H$ is nonzero. This is the case if symmetries constrain the DM vector to lie along a particular axis, for any magnetic field direction away from that special axis (or for a powder average). 
 \footnote{Case (2) can also occur if symmetries constrain the DM vector to lie in a plane and the magnetic field is not oriented perpendicular to this plane.}
 For example, this can occur when there is a mirror symmetry consisting of reflection across the 2D plane: this symmetry constrains $\vec D$ to lie perpendicular to the plane. 
 Even if the crystal does not have such a symmetry microscopically, it may be reasonable to expect that, within a 2D magnetic layer, the magnetic exchanges that develop under RG will depend only on the single layer, and that such an mirror symmetry may emerge as an approximate symmetry under RG. In particular, within an approximation where each magnetic layer is treated independently as a purely 2D system, a given SDRG step will involve only the configuration of valence bonds within the 2D layer that was generated in previous RG steps, and reflection across the plane becomes a symmetry.  In this case, only one component $D_i$ is relevant, and we find  $q=1$.
 
 Case (3): both $D_1,D_2$ are nonzero. This is the case if no symmetries constrain the DM vector. (This is also the case if symmetries constrain the DM vector to lie in a plane, and the magnetic field is oriented perpendicular to this plane.) Then both $D_1, D_2$ are relevant, and we find  $q=2$. 

Thus we find a different scaling form, with different powers of $T/H$, depending on the crystal or emergent symmetries as well as the direction of the magnetic field. 
 The linear scaling of 1T-TaS$_2$ ($q=0$) may be understood by noting that the  inversion centers in the 1T-TaS$_2$ crystal structure forbid DM interactions from being generated not only for nearest-neighbor bonds but also for 2nd, 3rd, 4th, etc neighbors, so it may be reasonable that no sizable DM interactions arise during RG. 
Quadratic  scaling  ($q=1$), as  in H$_3$LiIr$_2$O$_6$ and LiZn$_2$Mo$_3$O$_8$, can arise for generic or powder-averaged field directions through an emergent approximate mirror symmetry consisting of reflections across the 2D layer, a reasonable scenario for layered magnets.
Cubic scaling  ($q=2$) may thus be expected to arise in valence bond materials with 3D lattices, such as Ba$_2$YMoO$_6$\cite{Bos2010}. Experimental observation of the $q=2$ case is left for future work. 

\subsection{Magnetization scaling and NMR $1/T_1$}
\label{app:magnetization}
It is instructive to contrast the heat capacity with other experimental quantities. Magnetization, in particular, does not access the same information on the entropy of the spectrum.
Magnetization and susceptibility scaling has been previously discussed for SO(3) random singlets in the context of doped semiconductors
\cite{Bhatt1986,Bhatt1986a,Bhatt1986b}, as well as theoretically\cite{Singh2010} in the context of observed susceptibility scaling from the dilute impurity spins of synthetic herbertsmithite\cite{Lee2010}; the scaling is $M[H,T] \sim   H^{1-\gamma}$  for   $T \ll H$ and $M[H,T] \sim  H T^{-\gamma}$  for   $H \ll T$. 
\footnote{Large spin clusters which may arise in the course of the RG flow can contribute additional Curie (free spin) terms to the magnetization and susceptibility scaling.}
Note that magnetization at $T \ll H$ is determined by counting all states with $J < H$: any information on the low energy states with $\Delta E < T$, which determine the specific heat scaling $C[H,T]$, is not captured by the magnetization or susceptibility at $T \ll H$. In particular  the presence of weak DM interactions does not modify these expressions for magnetization or susceptibility scaling. 

Interestingly however, Maxwell's relation between entropy and magnetization gives an additional subdominant correction to the magnetization scaling at $T \ll H$; this correction encapsulates the heat capacity and does involve the $q$ exponent. Observe that the Maxwell relation $\pd S/ \pd H = \pd M/ \pd T$ (required by a thermodynamic potential $G[H,T]$ with curl[grad[$G[H,T]]]=0$) conveys information about the entropy to the temperature derivative of magnetization. 
 At $T \ll H$ the conventional magnetization scaling is expressed as $M \sim H^{1-\gamma}$ giving $\pd M / \pd T = 0$. 
 Maxwell's relation is satisfied by adding a subdominant negative term to the $T \ll H$ magnetization scaling, yielding the scaling function
\begin{align}
\label{eq:magnetizationscaling}
\frac{M[H,T]}{H} \ \sim& \ \ \frac{1}{H^{\gamma}} \ F^M_q\left[\frac{T}{H}\right]
\\
 F^M_q[X] \ \sim& \  \left\{ \begin{array}{cc}
 1 - m_1 X^{q+2}   & \quad X \ll 1 \\
 X^{-\gamma} & \quad X \gg 1 
 \end{array}
 \right.
 \nonumber
\end{align}
 Since we are in the $T \ll H$ regime, this additional $(T/H)^{q+2}$ correction is quite small. 
This subdominant magnetization scaling coefficient is given by 
 $m_1  = (q+\gamma)/((q+1)(q+2))((C/M) \left(H/T\right)^{q+1})[T{=}0]$ when heat capacity $C \sim H^{-q-\gamma} T^{q+1}$.
This relation  between the entropy measured in $C[H,T]$ and the magnetization $M[H,T]$ is also visible in the experimental data of Ref.~\onlinecite{Takagi2018}. There magnetization was found to obey a scaling relation that appears to be roughly captured by
 $M \sim H T^{-1/2} $ for $ H \ll T $, and,  for $T \ll H$, $
M \sim H^{1/2} $ with an additional subdominant correction consistent with a $(T/H)^3$ term added with a negative sign.
 Taking $\gamma \approx 1/2$ as extracted independently from the scaling of specific heat at zero field, these measurements are consistent with the theoretical expectations.
 
The scaling of heat capacity, equivalent to the scaling of the entropy, is also closely related to the scaling of $1/T_1 T$ where $T_1$ is the NMR spin-lattice relaxation time. The quantity $(T_1 T)^{-1}$ is related to the momentum-averaged (local) imaginary part of the spin susceptibility. Thus it serves as a measure of the spin-carrying states in the spectrum. 
NMR $1/T_1$ measurements complement the heat capacity by providing the information that the gapless spectrum carries spin excitations --- nontrivial information for weak SOC, which is of course satisfied by the spectrum associated with breaking valence bonds.

Note however that the broad distribution of couplings inherently implies that the spin-lattice relaxation, measured by the NMR lifetime $T_1$, should be described by a highly stretched exponential. A fit to an exponential with a single lifetime $T_1$ will likely not be appropriate. (When considering a single lifetime $T_1$, it is worth noting that the NMR Korringa law is a fine tuned case appropriate only for a metal: more generally the $1/T_1$ is linear in the density of states, $1/T_1 \sim C/\Gamma$ with some lifetime $\Gamma$ which is a priori not related to the density of states, and not known.)
In general there may be two contributions to the spin-lattice relaxation: one from the minority of spins that participate in the random-singlet scaling, and another from the phase of the bulk of the sites.
 However generally the fluctuations will be dominated by the few sites  in the tail of the energy distribution, i.e.\ the minority of spins in the random-singlet regime, producing a continuum of lifetimes for the spin-lattice relaxation and rendering exponential decay fits with a single lifetime $T_1$ difficult to interpret.

}

\bibliography{ScalingRefs}

\end{document}